\documentclass{article}
\usepackage[T1]{fontenc} 
\usepackage[utf8]{inputenc} 
\usepackage{ismir,amsmath,cite,url}
\usepackage{amsfonts}
\usepackage{amssymb}
\usepackage{calligra}
\usepackage{graphicx}
\usepackage{color}
\usepackage{algorithm}
\usepackage{enumitem}
\usepackage{caption}
\usepackage{subcaption}
\usepackage[noend]{algpseudocode}
\algnewcommand\algorithmicforeach{\textbf{for each}}
\algdef{S}[FOR]{ForEach}[1]{\algorithmicforeach\ #1\ \algorithmicdo}

\usepackage[bookmarks=false,hidelinks]{hyperref}

\usepackage[nolist]{acronym}
\begin{acronym}
    \acro{MIR}{Music Information Retrieval}
    \acro{MeanAP}{Mean Average Precision}
    \acro{MAP}{Maximum A Posteriori}
    \acro{ML}{Maximum Likelihood}
    \acro{AUC}{Area Under the receiver operating characteristic Curve}
    \acro{KB}{Knowledge-Based}
\end{acronym}

\title{Leveraging knowledge bases and parallel annotations for music genre translation}

\threeauthors
  {Elena V. Epure}{Deezer R\&D\\{}}
  {Anis Khlif} {Deezer R\&D \\ {\tt research@deezer.com}}
  {Romain Hennequin}{Deezer R\&D\\{}}

\begin{document}
\maketitle
\begin{abstract}
Prevalent efforts have been put in automatically inferring genres of musical items. Yet, the propose solutions often rely on simplifications and fail to address the diversity and subjectivity of music genres. Accounting for these has, though, many benefits for aligning knowledge sources, integrating data and enriching musical items with tags. 
Here, we choose a new angle for the genre study by seeking to predict what would be the genres of musical items in a target tag system, knowing the genres assigned to them within source tag systems. We call this a translation task and identify three cases: 1) no common annotated corpus between source and target tag systems exists, 2) such a large corpus exists, 3) only few common annotations exist. We propose the related solutions: a knowledge-based translation modeled as taxonomy mapping, a statistical translation modeled with maximum likelihood logistic regression; a hybrid translation modeled with maximum a posteriori logistic regression with priors given by the knowledge-based translation. During evaluation, the solutions fit well the identified cases and the hybrid translation is systematically the most effective w.r.t. multilabel  classification metrics. This is a first attempt to unify genre tag systems by leveraging both representation and interpretation diversity.
\end{abstract}
\section{Introduction}
\label{sec:introduction}
Music genres have been long studied as semantic dimensions of artists and tracks \cite{brackett2016categorizing}. 
Rooted in musicology, music experts have mainly undertaken this endeavour.
With digitization of music and prevalence of Internet music consumption, online communities have also shown increasing interest in annotating musical items with genres (e.g. creating folksonomies such as Lastfm).
In addition, crowd-sourced, web-based encyclopedias that describe and structure music-related knowledge including genres, have been created and openly disseminated \cite{auer2007dbpedia,swartz2002musicbrainz, Wang2010}.

Apart from ontologically describing musical items, genres are also among the most common attributes of tracks, albums and artists to which the users of music streaming services relate \cite{Mandel2010}.
Users resort to genres to discover music, create playlists, define their profiles, foster interactions with other users, etc.
Hence, being able to correctly infer music genres as metadata is central to such tasks.

Music genre is a challenging concept to model and highly subjective. 
Past studies \cite{Craft,Sordo2008,Lee,sturm2013} convey how difficult it is to agree upon shared definitions and interpretations, even for popular genres.
People interpret genres differently, influenced by their culture, personal preferences or acquired musicological knowledge \cite{Craft,Lee,Sordo2008}.
Genre representations within tag systems vary \cite{Schreiber2016} with respect to: 
the level of detail (how specialized genres can get); 
the coverage (which genres are considered); 
the genre interpretation (\emph{pop/rock} could be distinctly defined and interpreted across sources);
how genres are related (\emph{blues rock} is a subgenre of \emph{rock}, but not of \emph{blues} in the MuMu dataset \cite{Oramas2017, Hennequin2018}).
Divergences also result from the spelling variability (e.g. \emph{alternative rock} vs. \emph{alt. rock}).


The research question we address in this work is:
given annotations with genre tag systems of multiple sources, how to infer the equivalent annotations within a target tag system?
We refer to this as a translation task, 
but we do not necessarily seek to translate tags between languages.

When relying only on the definition of the sources and target tag systems, this task could be solved using taxonomy mapping \cite{AANEN2015,Prytkova:2015}.
A taxonomy is a classification schema with concepts organized from general to specialized.
The goal of taxonomy mapping is to align the concepts of the source and target taxonomies.
Related works integrate commercial catalogues \cite{Papadimitriou, AANEN2015}, align multi-lingual taxonomies \cite{Spohr2011, Wu:2016:CTA, Speer2017ConceptNet5A}
or restructure existing taxonomies \cite{Ponzetto:2009,Swoboda:2016, Prytkova:2015} in supervised or unsupervised manners. 
Ontology mapping \cite{otero2015ontology} is a similar task, in which additional relation properties and axioms can be exploited.

A solution focused on taxonomy mapping is nonetheless incomplete as it does not consider the application of the taxonomies in practice, which could reveal divergences in genre interpretation. 
Thus, we hypothesize that a robust translation is built not only on the definitions of genre tag systems, but also on their use for annotations. In accordance with the terminology of the Automatic Machine Translation domain \cite{conneau2017word}, we call a corpus of items jointly annotated by multiple sources a parallel corpus.


The contribution of the current work is a translation system that effectively leverages knowledge-based and statistical methods for genre translation in three cases: 
\begin{enumerate}[topsep=0pt,itemsep=-1ex,partopsep=1ex,parsep=1ex,leftmargin=*]
\item  A cold-start case, when genre tag systems of the target and sources are known, but there is no parallel corpus. 
We address this case with a \ac{KB} system based on taxonomy mapping (Section \ref{sec:knowledgeBasedMapping}).
\item Many parallel annotations are available allowing to learn mappings between genre interpretations (e.g. when some sources use \emph{alternative rock} the target tends to use \emph{alt. rock} and \emph{indie rock}).
To deal with this case, we use a simple linear multilabel classifier, namely a logistic regression model trained with \ac{ML} 
(Section \ref{sec:coocurenceBasedMapping}). 
\item The case in-between when less annotations are available and some target tags may be missing in the parallel corpus. We tackle this scenario with an hybrid Bayesian approach that leverages the \ac{KB} translation as a prior for the logistic regression model trained with \ac{MAP}.
This case, presented in Section \ref{sec:combiningStrategies}, is the most general.
Finding an effective solution for it has multiple positive implications for practice.
\end{enumerate}
We release the code of these methods for reproducibility\footnote{ \href{https://github.com/deezer/MusicGenreTranslation}{https://github.com/deezer/MusicGenreTranslation}}.

The \ac{MIR} community has extensively studied
the automatic genre annotation of musical items by exploiting the content (e.g. audio, lyrics) \cite{Oramas2017,Hennequin2018,Coviello2011}.
Other genre representations, tackled in \cite{Lisena2018,Achichi2018,Schreiber2016,Wang2010}, create genre graphs from multiple knowledge sources.
Yet, to our knowledge, there is no past work translating music genres from one tag system to another (e.g. from Discogs to Wikipedia) by leveraging the diversity of both genre representations and interpretations.

We resort to item annotation to assess the proposed translation methods.
To reflect a real-life context \cite{Bogdanov2017}, we consider a musical item annotated with multiple source tag systems; having multiple labels and not only broad genres such as \emph{rock}, but also very detailed subgenres, which results in predicting among hundreds of possibilities.
Lastly, combining multiple tag predictors in a Bayesian framework was done before \cite{Coviello2011,Tomasik2009}.
However, these works aggregate information from different predictors in the same tag system while we consider several tag systems.



\section{Notations and problem formulation}
\label{sec:problemFormulation}
In this work, we denote matrices 
by bold capital letters, $\textbf{M}$; vectors by bold lower case letters, $\textbf{v}$;
the $n$-th row vector of matrix $\textbf{M}$ by $\textbf{m}_n$;
scalars by italic lower case letters, $x$;
the coefficient at row $i$ and column $j$ of matrix $\textbf{M}$ by $m_{ij}$; the $i$-th element of vector $\textbf{v}$ by $v_{i}$.
Calligraphic font is used for sets of sets (e.g. $\mathcal{S}$) and capital letters for sets (e.g. $S$).

Let $\mathcal{D}$ be a set of tag systems, $\mathcal{S}$ a subset of $\mathcal{D}$, henceforth referred to as source tag systems, and $T \in \mathcal{D}, T \not\in \mathcal{S}$ henceforth referred to as a target tag system.
Further, we refer to a tag system as a tag set, but we stress that 
it may contain broader information such as relations between genres (e.g. taxonomies or ontologies). 
The research problem we address is:
given $\mathcal{S}, T$ and a set of tag annotations (e.g. associated with a given musical item) taken from $\mathcal{S}$, what would have been the corresponding tag annotations if the tags had been taken from $T$. 
We note $S = \cup_{E\in\mathcal{S}}E$ the union of the source tag systems, and $|S|$ its cardinality.

The approach we adopt consists in defining a translation scoring function $f: \mathcal{P}(S) \rightarrow \mathbb{R}^{|T|}$, where $\mathcal{P}$ denotes partitions over $S$, that predicts translation scores for every target tag from a set of source tags. 
Estimating such a scoring function is a standard setting for multilabel classification.


\section{Knowledge-based genre translation}
\label{sec:knowledgeBasedMapping}
We propose a translation method based on multiple genre taxonomies brought together under a genre graph. 
Section \ref{sec:genreKnowledgeRepresentation} introduces the graph types of concepts and relations and presents the genre taxonomies. 
In Section \ref{sec:normtags}, we show how we create the links between the genre taxonomies using advanced normalization and tokenization. 
In Section \ref{sec:learningMappingsWithDbpediaDisambiguation}, we define the translation scoring function $f$ by exploiting the genre graph structure and its relations.



\subsection{Building a knowledge-based genre graph}
\label{sec:genreKnowledgeRepresentation}

We automatically derive an undirected genre graph by aggregating multiple genre tag systems (e.g. taxonomies, ontologies or social tags), created by either experts or non-experts as in \cite{Speer2017ConceptNet5A, Schreiber2016,Lisena2018,Achichi2018}.
Its modular design allows to easily integrate new sources through a normalization pipeline that addresses much more variability of genre strings than the existing works \cite{Schreiber2015,Wang2010} (presented in Section \ref{sec:normtags}).
The knowledge sources used to build the current version of the genre graph are: DBpedia (English, 12443 genres), and 
Lastfm (327 genres), Tagtraum (296 genres) and Discogs (296 genres)--the taxonomies released in the 2018 MediaVal AcousticBrainz Genre Task \cite{Bogdanov2017}. 
The Discogs genre taxonomy is pre-defined by experts. 
The Lastfm and Tagtraum genre taxonomies are automatically inferred from social tags with the approach proposed by Schreiber \cite{Schreiber2015},
followed by a manual processing \cite{Bogdanov2017}.

The types of relations between genres vary across sources. 
In DBpedia, the retrieved types for each genre are: subgenres, origins, aliases--various spellings of the same genre, and derivatives--genres which are influenced by this genre, but could not be considered subgenres. 
The other knowledge sources contain only subgenre relations.

\begin{figure}[h]
\centering
\includegraphics[width=0.45\textwidth]{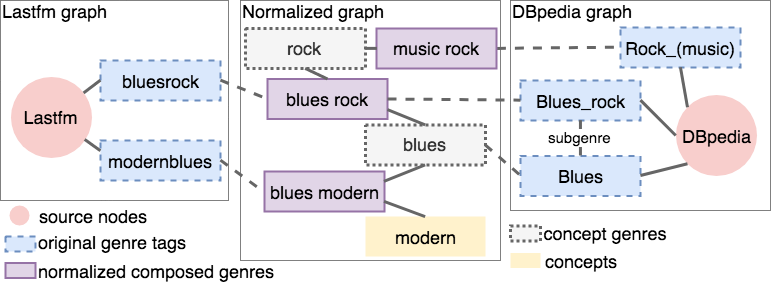}
\caption{Genre graph extract. Dashed edges link the original genre tags to their normalized versions.}
\label{fig:genregraph}
\end{figure}

Each genre tag system becomes a graph by adding a source node that connects all the genre tags as in Figure \ref{fig:genregraph}.
Then, to connect these decentralized graphs, a normalized graph is produced from all available tags.
Each original tag is connected to its normalized form in the normalized graph.
The description of how we normalize genres and create the normalized graph is continued in Section \ref{sec:normtags}.

\subsection{Normalizing genre tags}
\label{sec:normtags}
We create a more robust normalization pipeline compared to the related works \cite{AANEN2015, Spohr2011,Schreiber2015,Wang2010} that, apart from basic tokenization and normalization, also separates words written together (e.g. \emph{poprock} in \emph{pop} and \emph{rock}).
The basic tokenization splits tags by non-alphanumeric characters (e.g. "-", "\_").
The basic normalization converts tags to lower case and brings tags containing "\&", "+" and "'n'" to the same form (e.g. \emph{d+b}, \emph{drum'n'bass} and \emph{drum and bass}).

For the advanced tokenization, we use a modified trie \cite{DeLaBriandais:1959} and a probabilistic tokenization built on Wikipedia unigrams \cite{word_ninja_2019}.
A trie is a tree data structure that efficiently stores and retrieves strings. 
Each node has a char and a flag to mark if the path from the root to it forms a word.
We modify the way we populate the trie as follows.
At first, we sort the tokens obtained from the basic tokenization and normalization, ascendingly  by length. 
Then, we add the tokens of DBpedia with less than $l$ letters directly to the trie\footnote{DBpedia seeds the trie as it has the highest coverage and we set $l$=7.}.
For the others, we attempt to split them using the trie and only the unknown words are added to the trie. 

The tokenization using the trie is a recursive greedy algorithm that aims at matching the longest possible words in the trie.
If a recursion fails, we explore the path with the next best previous word instead. 
If we assess the split output as incorrect, meaning that it results in too many short words, in short suffixes, or fails to split a large tag, then we use the probabilistic tokenization.

The probabilistic tokenization uses dynamic programming to find the words best maximizing their probability product.
The frequency of each word, assuming that they are independently distributed, is approximated using the Zipf's law \cite{Zipf1949} to $ \frac{1}{nlog(N)}$, where $n$ is the word rank \cite{word_rank_2019} and $N$ is the total number of Wikipedia unigrams.
We again assess the split output.
Some extra conditions are added besides those presented in the previous paragraph:
a Wikipedia split is incorrect if there are single letters as middle words and if no word is already contained in the trie\footnote{As we already added to the trie short genre and concept tags from multiple sources, we assume the probability of all words to be new is low.}.
If this tokenzation fails, we add the token as it is.

Finally, we transform the obtained tokens in nodes in the normalized graph (see Figure \ref{fig:genregraph}). 
There are three types of nodes: 
1) normalized composed genres (e.g. \emph{altern rock},  \emph{deep house}), 
2) concepts which are words that do not represent genres but are part of the name of multiple genres (e.g. \emph{nu} in \emph{nu jazz} and \emph{nu metal}); 3) concept genres which are standalone genres but can be also part of composed genres (e.g. \emph{punk} in \emph{post punk}).
If a genre is tokenized, its tokens are sorted and concatenated becoming a composed genre node as in \cite{Schreiber2015} (e.g. \emph{music rock} in Figure \ref{fig:genregraph}).
This node is then connected to its concept and concept genre nodes. 

\subsection{Translating Genres through DBpedia Mapping}
\label{sec:learningMappingsWithDbpediaDisambiguation}
Using intermediate mapping spaces such as taxonomies or pivot languages has been explored in past works to match multi-lingual \cite{Spohr2011,Wu:2016:CTA}, multi-cultural \cite{Prytkova:2015} or e-commerce \cite{AANEN2015, Ponzetto:2009} taxonomies.
Similar to \cite{Prytkova:2015}, we use DBpedia, an ontology derived from Wikipedia infoboxes \cite{Lehmann2015} as it has the highest genre coverage and quite high quality. 
However, to map a genre to DBpedia genres, we avoid using string similarity as it can be very noisy (e.g. \emph{pop} vs. \emph{bop}).
Instead, we leverage genre knowledge to create a mapping strategy as we further present.
Most related works rely on the structure of taxonomies for mapping the source and target concepts \cite{Spohr2011,Papadimitriou,AANEN2015,Prytkova:2015,Wu:2016:CTA}.
Our solution uses structural information too, but differently. 
Specifically, we use the neighbours of the source and target concepts and the structure of the directed DBpedia graph.


We map each genre of the source and target tag system, to the genres of the DBpedia ontology: $B \in \mathcal{D}$.
We assume $B \notin \mathcal{S}$ and $B \neq T$.
For each input tag system $D$, with $D \in \mathcal{S}$ or $D=T$, the output of the mapping is a matrix $\textbf{Z}^{D} \in \mathbb{R}^{|D| \times |B|}$, where each row represents the relatedness of a genre tag from $D$ to the DBpedia genres.
We compute the mapping matrix $\textbf{Z}^{D}$ by applying the following steps for each tag $s\in D$:

\begin{enumerate}[topsep=0pt,itemsep=-1ex,partopsep=1ex,parsep=1ex,leftmargin=*]
    \item Normalize $s$ with the process described in Section \ref{sec:normtags} (\emph{e.g.} \emph{Rock/Pop} becomes \emph{pop rock}).

    \item Check if the normalized $s$ equals any normalized genre of $B$.
    If true, all entries in $\textbf{Z}_{D}$ linked to the DBpedia aliases of the found genres are set to $1$ and all others to $0$ (e.g. \emph{acid house} is mapped to \emph{Acid\_house}, with aliases \emph{Acid\_(electronic\_music)}, \emph{Warehouse\_music}, etc.).

    \item If the normalized $s$ is not in $B$, then map it using its context genres in $D$: compound $s$ with each parent tag in $D$ and check if the normalized compounded tag equals any normalized genre of $B$ (inspired from \cite{Wu:2016:CTA}). If true, proceed as in Step 1. (e.g. \emph{stoner} has parent \emph{rock} in Lastfm; search by \emph{rock stoner} and map it to \emph{Stoner\_rock}).

    \item If Steps 2 and 3 are unsuccessful, consider two cases:
    \begin{enumerate}[topsep=0pt,itemsep=-1ex,partopsep=1ex,parsep=1ex,leftmargin=*]
        \item $s$ is a concept genre as defined in Section \ref{sec:normtags}. First, retrieve the DBpedia directed subgraph composed of the nodes which contain the normalized $s$ as a substring in their normalized form. 
        Second, map $s$ to the nodes with the highest in-degree centrality \cite{Bang-Jensen2008} in this subgraph.
        The intuition is that concept genre nodes are more likely fundamental music genres; hence they tend to have many subgenres or related genres.
        Third, assign to the selected DBpedia genres and their aliases a score of $1$ divided by the number of selected nodes, and to the others $0$ (e.g. \emph{rock} does not exist as is in DBpedia. 
        To map it, we retrieve all tags that contain it such as \emph{Punk\_rock}, \emph{Art\_rock}, \emph{Rock\_music}, etc. 
        We observe that \emph{Rock\_music} is the most connected node in the subgraph with the genres containing \emph{rock}. As only one node is selected, we assign to it and its aliases a score of $1$).
        
        \item $s$ is a composed genre as defined in Section \ref{sec:normtags}. First, select from the normalized genres in $B$ those that share the greatest number of words with $s$. 
        Second, select from this list, the genres with the highest number of shared concept genres--if it is $0$, then the initial selection is kept unchanged. 
        Third, assign scores as in Step 4(a). 
    \end{enumerate}
\item For each genre in $B$ associated to $s$ in Steps 1--4, propagate half of the value of its score to its neighbors in $B$.
The intuition is that parent genres or subgenres could be relevant and sometimes specified by other sources.
\end{enumerate}
For each $s$ not mapped in the previous process, we compute its scores by averaging the rows in $\textbf{Z}^{D}$ of its related genres in the input taxonomy $D$ (e.g. for \emph{aor} which is not found in DBpedia, we compute the scores by assigning it the scores obtained for \emph{rock}, its parent genre in Discogs).
Finally, the relatedness of a source genre $s \in S$ and a target $t \in T$ is computed using cosine similarity between their corresponding rows $\textbf{s}=\textbf{z}^S_s$ and $\textbf{t}=\textbf{z}^T_t$ in the mapping matrices, $\textbf{Z}^S$ and $\textbf{Z}^T$. We define $\textbf{W}^{KB} \in\mathbb{R}^{|T| \times |S|}$ such that $w^{KB}_{ts} = \frac{\textbf{s}^T\textbf{t}}{||\textbf{s}||_2 ||\textbf{t}||_2}$. The translation scoring function is:
\begin{equation}
f_t(\{s_{1}, s_{2}, \dots, s_K\}) = \sum_{k=1}^{K}w^{KB}_{ts_{k}} = \textbf{x}^{T}\textbf{w}^{KB}_{t},
\end{equation}
where $\textbf{x}$ is the binary encoded vector of $\{s_{1},\dots,s_K\}$.

\section{Data-informed genre translation}
In this section, we consider that a parallel corpus is available and present two statistical approaches: \ac{ML} that relies only on annotations (Section \ref{sec:coocurenceBasedMapping}), and \ac{MAP} that leverages the \ac{KB} results
as a prior knowledge
(Section \ref{sec:combiningStrategies}).

\label{sec:dataInformedGenreTranslation}
\subsection{
Maximum Likelihood logistic regression}
\label{sec:coocurenceBasedMapping}
In statistical approaches to the tag translation task, we seek to train a parametric mapping to model the probability $P(\textbf{y} | \textbf{x})$ of having a collection of target tags (encoded as a binary vector $\textbf{y} \in \{0,1\}^{|T|}$) given the source tags (encoded as a binary vector $\textbf{x} \in \{0,1\}^{|S|}$). We assume the independence of the target tags, and only seek to model the conditional probabilities $P(y_{t} | \textbf{x})$. This comes down to training $|T|$ binary classifiers, also known as binary relevance.
There are more elaborated settings for doing multilabel classification without the target tag independence assumption.
We notably also tested classifiers chain \cite{read2011classifier}, but it did not result in significant improvement over the results presented in Section \ref{sec:results}, while increasing the system complexity. 
We propose to implement binary relevance with logistic regression \cite{Walker67}. Logistic regression models the probability of having the $t$-th target tag $y_{t}$ given the source tags $\textbf{x}$ and the parameters of the logistic regression $\theta = \{\textbf{W}, \textbf{b} \}$, $\textbf{W} \in \mathbb{R}^{|T| \times |S|}$ $\textbf{b} \in \mathbb{R}^{|T|}$; as:
\begin{equation} \label{logreg:1}
    P(y_{t} = 1 | \textbf{x}, \theta) = \sigma(\textbf{w}_{t}^{T}\textbf{x} + b_{t})
\end{equation}
\noindent where 
$\sigma(x)=\frac{1}{1+\exp(-x)}$. $\textbf{W}$ is called the weights matrix and $\textbf{b}$ the bias.
\noindent 
Note that, for the statistical approaches, the scoring function $f$ introduced in Section \ref{sec:problemFormulation} is defined here as ${f(\{s_1,s_2,\ldots,s_K\})} = {\left(P(y_1=1|\textbf{x},\theta),\ldots,P(y_{|T|}=1|\textbf{x},\theta)\right)}$. 
To train a logistic regression model we maximize the log-likelihood of the targets, given the source tags, w.r.t. the parameters $\theta$:
\begin{equation} \label{logreg}
\begin{split}
    \mathcal{L} = \log P(\textbf{Y} | \textbf{X}; \theta) & = \sum_{n=1}^{N} \textbf{y}_{n}^{T}\log(\sigma(\textbf{W}\textbf{x}_{n} + \textbf{b})) + \\
    &(\textbf{1}-\textbf{y}_{n})^{T}\log(\textbf{1}-\sigma(\textbf{W}\textbf{x}_{n} + \textbf{b}))
\end{split}
\end{equation}
where $N$ is the size of the parallel corpus; $\textbf{X} = [\textbf{x}_{1},..., \textbf{x}_{N}]^T \in \{0,1\}^{N \times |S|}$; and
$\textbf{Y} = [\textbf{y}_{1},..., \textbf{y}_{N}]^T \in \{0,1\}^{N \times |T|}$. In practice the regularization term $\frac{1}{2}||\textbf{W}||_{F}^{2}$ is added to $\mathcal{L}$ in the objective, where $||\cdot||_{F}$ denotes the Frobenius norm on matrices, to limit overfitting.

\subsection{A unified translation model}
\label{sec:combiningStrategies}
While \ac{ML} logistic regression can be expected to work well with large amounts of parallel annotations, 
they will not adapt well to settings where no or little parallel data is available. In a real-life scenario, the size of the parallel corpus can range from zero to tens of thousands of samples, which precludes systematically favoring one or the other. Defining a criterion for when to switch from \ac{KB} to statistical translation is arduous since this criterion would depend on the number of source and target tags as well as on their distribution. Ideally, we would like to have knowledge-based performances when no parallel data is available, and a smooth way to transition towards more data-abundant settings.
This leads us to consider the translation table $\textbf{W}^{KB}$ given by the \ac{KB} system as a prior
in a Bayesian framework, using the \ac{MAP} \cite{bishop2006pattern} objective. Instead of maximizing the likelihood of the target tags, given source tags and parameters, we maximize the posterior probability of the parameters given the source and target tags:
\begin{equation} \label{bayes}
    P(\theta | \textbf{x}, \textbf{y}) \propto P(\textbf{y} | \textbf{x}; \theta)P(\theta| \textbf{x}) = P(\textbf{y} | \textbf{x}; \theta)P(\theta).
\end{equation}
\noindent By assuming, for each target tag $t$ a normal distribution for $\textbf{w}_{t}$ centered around $\textbf{w}^{KB}_t$ with a precision matrix ${\bm \Lambda} = \lambda^2 \textbf{I}$ ($\lambda$ is independent of $t$), we can write the logarithm of the prior distribution as:
\begin{equation} \label{prior:1}
    \log(P(\theta)) = \frac{\lambda^{2}}{2} ||\textbf{W} - \textbf{W}^{KB}||^{2} + \text{cte.}
\end{equation}

\noindent We also consider 
a centered Gaussian prior on the bias (corresponding to a $l^2$ regularization). We then define:
\begin{equation} \label{prior:2}
    \mathcal{L}_{\text{prior}} = \frac{\lambda^{2}}{2} ||\textbf{W} - \textbf{W}^{KB}||^{2} + \nu ||\textbf{b}||^2.
\end{equation}

\noindent Using 
\eqref{logreg}, \eqref{bayes} and \eqref{prior:2}, the final MAP objective 
becomes $\mathcal{L}_{\text{MAP}} = \mathcal{L} + \mathcal{L}_{\text{prior}}$, where the first term is the loss of Eqn \eqref{logreg}, and the second can be seen as a regularization term on the weight matrix $\textbf{W}$ that penalizes its straying away from the priors. 
$\mathcal{L}$ depends on the number of training samples, while $\mathcal{L}_{\text{prior}}$ does not. Therefore, $\mathcal{L}$ becomes the predominant term in the loss as the size of the training data grows, leading to an objective function very close to the one of the logistic regression of Section \ref{sec:coocurenceBasedMapping}. Conversely, when little data is available, we can expect the performances to be close or better than those of the \ac{KB} system.


When a large parallel corpus is available, we can choose $\lambda$
with grid search on a validation set. This is computationally expensive, and does not adapt well
when the parallel corpus is small. For the sake of adaptability, we hereby propose a principled way inspired by \cite{gelman2008weakly} to choose $\lambda$, that does not require a lot of data while achieving top results. The rationale builds on the limited effective range of the logistic regression parameters. A shift of 5 of $w_{ts}$ in the logit scale can move the probability associated with the target tag $t$ from 0.5 to 0.99 or from 0.01 to 0.5. 
Hence, we would tend to choose $\lambda$ in such a way that bigger shifts in the predicted probability of the target tag, which is the result of the added shifts for each annotated source tag, are unlikely. If we note $\bar{N}_{S}$ the average number of source tags per sample (which can be estimated with a few samples), this would mean restricting the coefficients $w_{ts}$ from shifting by more than $\frac{5}{\bar{N}_{S}}$.  For a normal variable $X \sim \mathcal{N}(\mu, \frac{1}{{{\lambda}}})$ we have $P(X \in [\mu-\frac{2}{\sqrt{\lambda}}, \mu+\frac{2}{\sqrt{\lambda}}]) \approx 95\%$, we therefore propose to choose precision $\lambda$ such that: \begin{equation} \label{eq:elicitation}
    \frac{2}{\sqrt{\lambda}}  \approx \frac{5}{\bar{N}_{S}}
\end{equation}

\section{Experiments}

We report the performances of the proposed models on a recording-based tag translation task. 
This also serves as an indirect evaluation of the DBpedia mapping, which, in a work dedicated to taxonomy mapping, could have been assessed by experts. 
Due to its novelty, we do not benchmark our work against other genre-related research from \ac{MIR}. 

\label{sec:experiments}
\subsection{Dataset}
\label{sec:dataset}
The dataset used in the experiments was created from the dataset used in the 2018 AcousticBrainz Genre Task, part of the MediaEval benchmarking initiative \cite{Bogdanov2017}.
The dataset in its original form was aimed at testing the automatic genre annotation from content-based features of musical items in a more challenging setup compared to past works.
For each item, annotations from different sources were available, each source taxonomy was much more detailed with hundreds of genres-subgenres, and the overall task was modeled as a multi-label classification.
The sources were already introduced in Section \ref{sec:genreKnowledgeRepresentation}.
We further describe how the provided dataset was created.
In Discogs, the release annotation was propagated to tracks.
In Lastfm and Tagtraum, each track was annotated with music genres and subgenres from the derived taxonomies \cite{Bogdanov2017}.
We present an overview of the dataset in Table \ref{datasets}.
{\small
\begin{table}[h]
\centering
\begin{tabular}{l|l|l|l}
Dataset & Discogs & Lastfm & Tagtraum \\
\hline
Annotation type & Expert & User & User \\
Items & 1,098,337 & 686,979 & 589,584\\
Number genres & 315 & 327 & 296\\
\hline
\end{tabular}
\caption{Description of the dataset \cite{Bogdanov2017}.}\label{datasets}
\end{table}
}

Although the data was already split between development, validation and test \cite{Bogdanov2017}, we brought several modifications to accommodate the translation task.
We created a large dataset comprising the original development and validation data.
In order to assess a notion of confidence on the computed metrics, we resorted to K-Fold cross validation \cite{friedman2001elements}.
For each possible target, we splitted the data in $4$ folds using stratified sampling. 
First, we filtered out the items which were not annotated in the target tag system.
Then, we used an altered version of the iterative stratification algorithm in \cite{Sechidis:2011} in order to ensure that the proportion of items for each target label was roughly the same across folds. 
Following \cite{Flexer2007ACL}, we added the constraint that items belonging to the same artist had to be assigned to the same fold.
For that, we used MusicBrainz artist ids retrieved from the recording ids provided in the MediaEval data.


\subsection{Evaluation setup}
\label{sec:evaluationSetup}

The presented models output a score for each target tag that relates to the confidence of this tag being used in the target annotation. 
We evaluated these outputs with a ranking metric called \ac{AUC}, as commonly done in multilabel classification \cite{Oramas2017}. The (\emph{macro}) averaging is over target tags and measures the ability of the system to rank higher a positive tag than a negative one. Specifically, shifting the values in a column by the same factor (or changing the values of $\textbf{b}$ in the logistic regression) does not change the \ac{AUC} macro score, being in that sense, unaffected by item popularity.

We evaluated the logistic regression models on each fold and trained on the three others. 
We uniformly subsampled the training data to simulate low data availability and chose subsampling factors as powers of $2$ 
between $2^{-13}$ and $1$. Consequently, for the smallest subsampling factors, some source and target tags may not be present in the training data. We used scikit-learn \cite{scikit-learn} implementation for \ac{ML} logistic regression, with L-BFGS as the solver. We wrote a Tensorflow \cite{tensorflow2015-whitepaper} implementation of \ac{MAP} logistic regression. The Adam optimizer \cite{kingma2014adam} was used, with a learning rate of $0.5$. 
We trained the model for $500$ epochs with batches of size $100000$ or with the full training set if there were less samples. $\lambda$ was chosen 
using Eqn \ref{eq:elicitation}.
\subsection{Results}
\label{sec:results}

\begin{figure}[htbp]
    \centering
    \begin{subfigure}[b]{0.45\textwidth} 
        \centering \includegraphics[width=\textwidth]{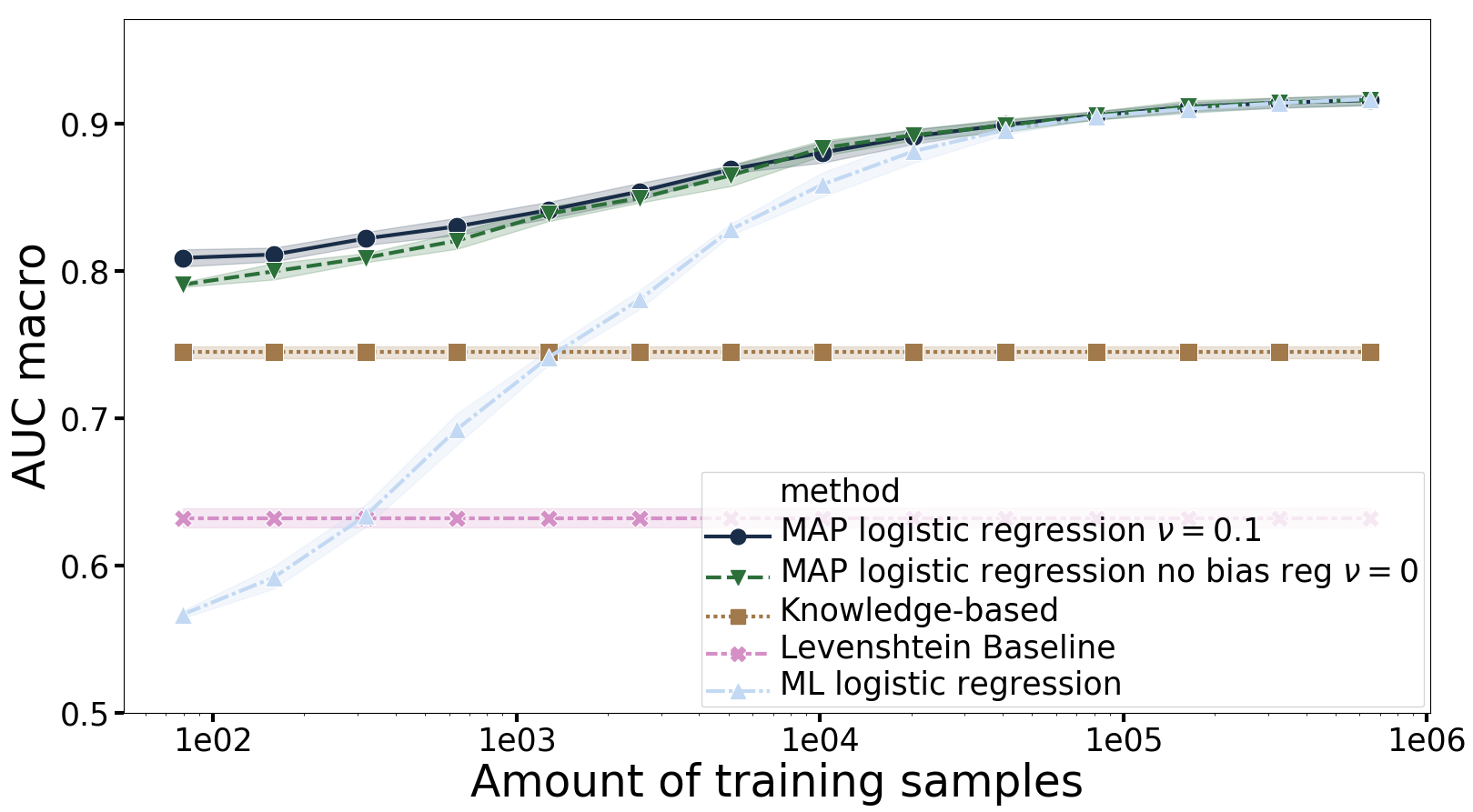}
        \caption{Translation from Lastfm and Tagtraum to Discogs}\label{fig:translation_to_discogs}
    \end{subfigure}

    \begin{subfigure}[b]{0.45\textwidth}
        \centering \includegraphics[width=\textwidth]{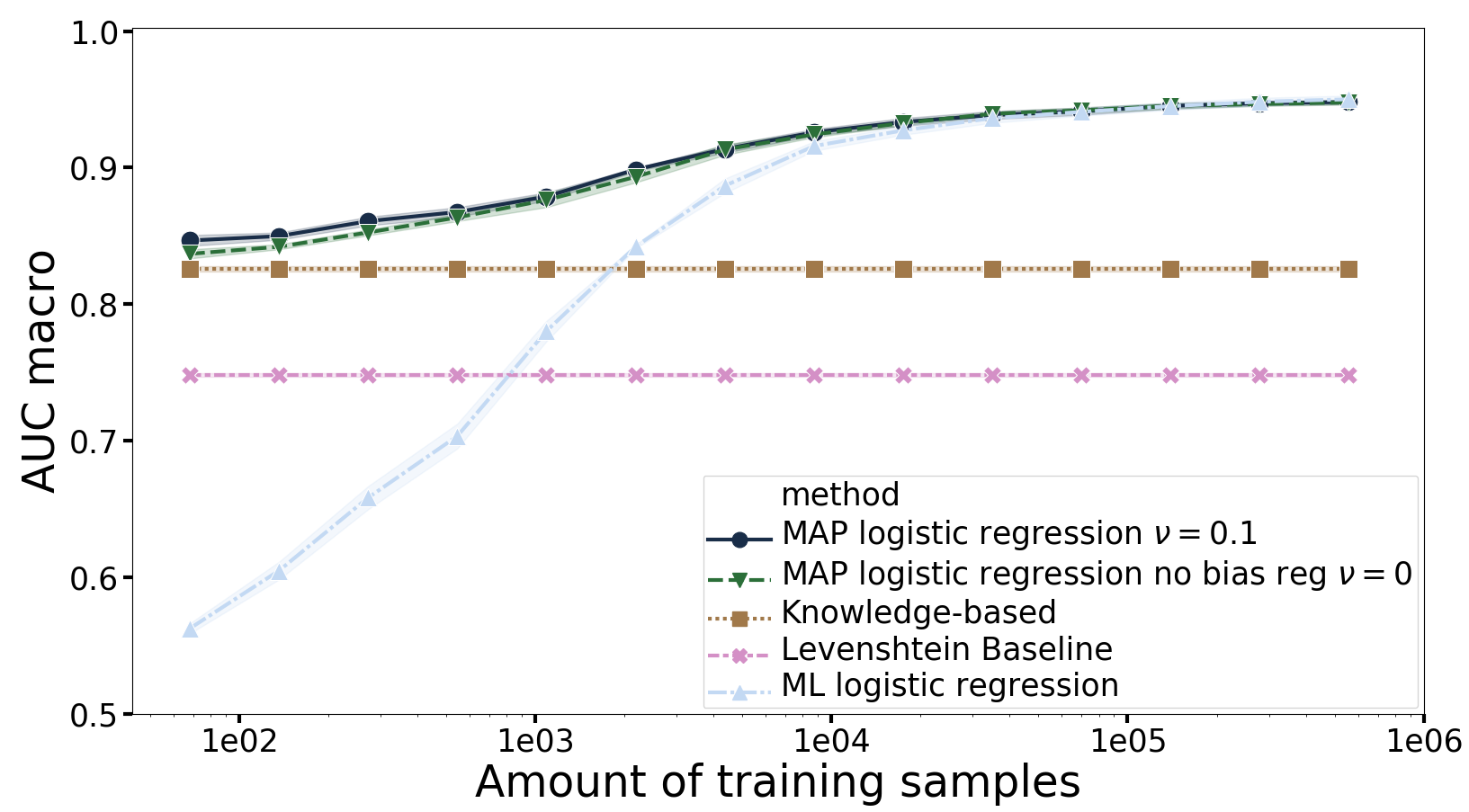}
        \caption{Translation from Discogs and Tagtraum to Lastfm}\label{fig:translation_to_lastfm}
    \end{subfigure}
    \caption{\ac{AUC} scores for the three models per amount of training samples (log-scale). The width of the area around the lines marks the standard deviation computed on folds. \ac{KB} yields a constant value as it is data-independent.}\label{fig:AUC_from_two_sources}
\end{figure}

Figure \ref{fig:AUC_from_two_sources} illustrates how the \ac{ML} translation 
eventually outperforms the \ac{KB} model when enough data is available, while the latter performs much better when little data is available.
The \ac{MAP} translation successfully builds on the \ac{KB} translation to yield the best results across the whole data availability spectrum.
A simple baseline based on tag Levenshtein distance is also shown.
Using only a source instead of two (e.g. only Lastfm) led to the same kind of behavior.
While we currently proposed one method to obtain the \ac{KB} translation table, we could also imagine it replaced by an expert-created one, if desired.

The fact that the \ac{MAP} logistic regression performs consistently well on all the translation tasks is favorable evidence towards the choice of $\lambda$ given in Eqn \ref{eq:elicitation} as a good default, which we also confirmed using a grid search. 
Furthermore, we see that the \ac{MAP} logistic regression leverages even low amounts of training data to improve over the \ac{KB} model, and even more so when applying regularization on the bias. We 
further explain this effect by analyzing how the \ac{AUC} scores compare on a per-tag basis.

\begin{figure}[h]
\centering
\includegraphics[width=0.43\textwidth]{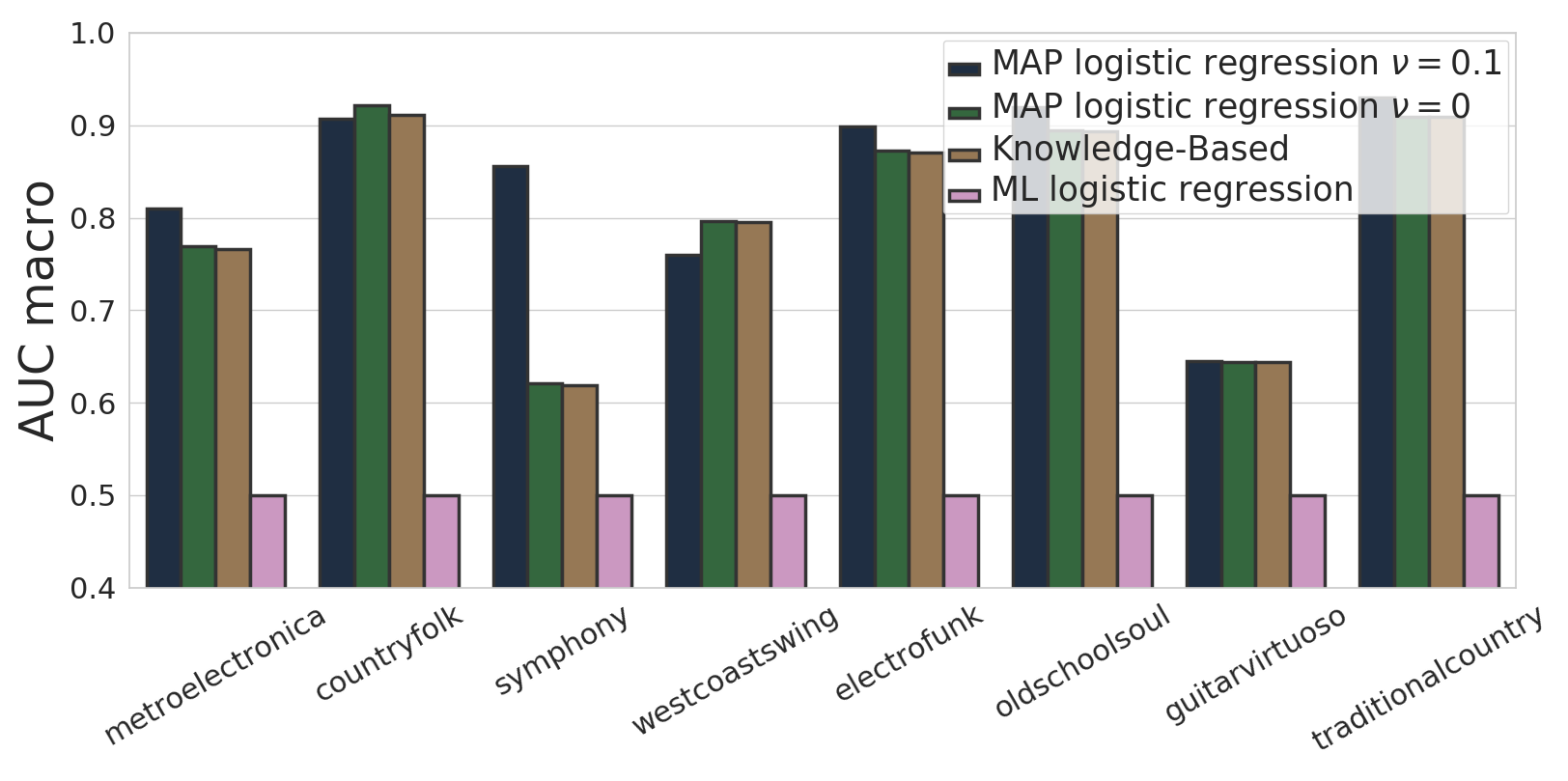}
\caption{\ac{AUC} scores for tags that were not in the training set,
subsampled by a factor of $2^{-12}$ with Lastfm as target.}
\label{fig:resultspermissingitems}
\end{figure}

Figure \ref{fig:AUC_from_two_sources} shows that MAP logistic regression with bias regularization can achieve better \ac{AUC} scores than the \ac{KB} translation, even on target tags absent during training. 
We argue that this is due to the regularization term on the bias that enables to learn negative correlations between tags. When no bias regularization is used, the optimal set of parameters for a tag $t$ missing from the training data is:  $({\textbf{w}}_{t}^{\star}, b_{t}^{\star}) = ({\textbf{w}}_{t}^{\text{KB}}, -\infty)$. 
Indeed, we see in Figure \ref{fig:resultspermissingitems} that the \ac{MAP} results are very close to those of the \ac{KB} model. This is not true anymore with bias regularization. The gradient of the cost function w.r.t. $\textbf{w}_{t}$ can be written as:
\begin{equation} \label{eq:gradmaplogreg}
    \frac{\partial \mathcal{L}_{\text{MAP}}}{\partial \textbf{w}_{t}} = [\hat{s}\sigma(w_{ts} + b_{t}) + \lambda^{2}(w_{ts} - w_{ts}^{\text{KB}})]_{1 \leq s \leq |\mathcal{S}|}
\end{equation}
\noindent where $\hat{s}$ is the number of times (possibly $0$) the source tag $s$ appears in the training set. We therefore see that as $w_{ts}$ gets closer to $w_{ts}^{KB}$, the term $\hat{s}\sigma(w_{ts} + b_{t})$ will start to outweigh the second. Applying a gradient step will tend to decrease $w_{ts}$, away from $w_{ts}^{KB}$, and even more so when $\hat{s}$ is large (popular tags) and $b_{t}$ is close to 0 (controlled by the regularization term), hence the negative correlation.

Finally, it is worthwhile to mention that the \ac{AUC} metric relies on occurrences and is thus arguably biased towards statistical methods. 
We end this section by taking a qualitative look at how statistics modified the similarities between source and target tags, in particular for those with very different \ac{KB} and \ac{ML} \ac{AUC} scores. These differences fall under four explanations:
\begin{itemize}[topsep=0pt,itemsep=-1ex,partopsep=1ex,parsep=1ex,leftmargin=*]
    \item Annotation noise: Statistical models learn a very high similarity between the Discogs tag \emph{italo-disco} and the Lastfm tag \emph{classicalbritishheavymetal}. Both indeed often co-occur in the data, but are ontologically unrelated.
    
    \item The target tag does not have a suitable equivalent in the source taxonomies. Some Latin and Caribbean music genres like \emph{cumbia}, \emph{fado}, \emph{rocksteady} or \emph{forró} are present in Discogs but are not in Lastfm or Tagtraum. Thus, the mapping to DBpedia, described in Section \ref{sec:learningMappingsWithDbpediaDisambiguation}, fails.
    
    \item The considered tag is highly ambiguous. Take the example of the tag \emph{classical}. Besides the identical counterparts, knowledge-based translation tables also indicates relatedness to some subgenres of \emph{jazz}. However, the specific translation task on which we evaluate appears to be more biased towards an understanding of \emph{classical} that relates to subgenres of \emph{metal} and \emph{electronic} music (\emph{symphonicmetal}, \emph{germanmetal}, \emph{postmodernelectronicpop}).
    
    \item The existing genre representations are incomplete or noisy. For instance, \emph{baroque} has a counterpart in each taxonomy, but no direct link with \emph{classical} in DBpedia. 
    Statistical models find high correlation in the data between those two tags and so achieve better \ac{AUC} scores.
\end{itemize}

\section{Conclusion}
\label{sec:conclusion}
In this work, we investigated the translation of tags from various source tag systems to a common target tag system.
We show that the availability of large amounts of data advantages statistical methods over the knowledge-based one in terms of multilabel classification metrics.
Moreover, the proposed hybrid method consistently outperforms both other methods on the whole range of data availability.


Although we did not address multi-language tag systems, 
both the knowledge-based approach that uses a mapping through the multilingual DBpedia, and the data-informed approach that only takes advantages of parallel annotations and is then insensitive to language, should be able to handle it. 
As future work, we aim to gather multilingual music genre datasets in order to confirm this claim.
We also aim to exploit more thoroughly the genre graph we created by adding more knowledge sources and generating genre representations as node embeddings.
We also consider modelling the tag annotation noise, such as missing or spurious tags, or tag bombing, in order to filter it out.

\bibliography{ISMIR2019}

\end{document}